\documentclass[aps,prl,twocolumn]{revtex4-1}  

\usepackage{bbm,graphicx,bm,bbm,amsmath,hyperref}   

\def\ave#1{\langle #1 \rangle}
\def\cL{{\cal L}}
\def\ii{{\rm i}}
\def\sx{\sigma^{\rm x}}
\def\sy{\sigma^{\rm y}}
\def\sz{\sigma^{\rm z}}
\def\s1{{\mathbbm{1}_2}}

\def\tr#1{{\rm tr}{{#1}}}

\def\etal#1{#1}

\def\tit#1{}

\begin{document}

\title{Exact large-deviation statistics for a nonequilibrium quantum spin chain}

\author{Marko \v Znidari\v c}
\affiliation{
Physics Department, Faculty of Mathematics and Physics, University of Ljubljana, Ljubljana, Slovenia}

\date{\today}

\begin{abstract}
We consider a one-dimensional XX spin chain in a nonequilibrium setting with a Lindblad-type boundary driving. By calculating large deviation rate function in the thermodynamic limit, being a generalization of free energy to a nonequilibrium setting, we obtain a complete distribution of current, including closed expressions for lower-order cumulants. We also identify two phase-transition-like behaviors in either the thermodynamic limit, at which the current probability distribution becomes discontinuous, or, at maximal driving, when the range of possible current values changes discontinuously. In the thermodynamic limit the current has a finite upper and lower bound. We also explicitly confirm nonequilibrium fluctuation relation and show that the current distribution is the same under mapping of the coupling strength $\Gamma \to 1/\Gamma$.
\end{abstract}



\maketitle

Systems out of equilibrium are of wide importance. On a practical side, all dynamical processes, for instance in devices performing some useful work, involve nonequilibrium states supporting nonzero currents. Also, phenomena not possible in equilibrium, for instance finite-temperature phase transitions in local one-dimensional systems, become possible out of equilibrium. Unfortunately, understanding nonequilibrium physics is more difficult than understanding equilibrium. For a start, there is no general nonequilibrium formalism analogous to equilibrium measures and thermodynamic potentials.

The best method available for study of stationary nonequilibrium systems is the so-called large deviation (LD) formalism~\cite{Oono:89,Derrida:07,Touchette}, which enables us to assess, in addition to most probable values, also the whole distribution function. Because the formal mathematical frame-set used in the LD approach is very similar to the one used in equilibrium statistical physics, one can argue that the functions provided by the LD method are generalizations of equilibrium thermodynamic potentials like the entropy and free energy. In the last 15 years LD formalism has been successfully applied to classical nonequilibrium systems, most notably to exclusion processes~\cite{Derrida:98,Derrida:01,Essler:11,Gorissen:12,Lazarescu:13}. A LD function of heat current in a chain of harmonic oscillators described by a quantum Langevin equation has been calculated in Ref.~\cite{Keiji:07}. For quantum systems much less is known because exact quantum solutions are typically more difficult to obtain. In fact, full LD formalism has been applied to quantum systems only very recently~\cite{Garrahan:10}, mainly to single or two qubit systems~\cite{Budini:10,Li:11,Lesanovsky:12}, or, within a mean field approximation also in an infinite Ising spin chain~\cite{Ates:12}. Because interesting phenomena, like phase transitions, are possible only in the thermodynamic limit (TDL) it is important to have solutions, possibly exact ones, for thermodynamically large many-body systems. No exact LD statistics are known for quantum systems in the TDL, and the goal of this work is to provide one. 

Hoping for an an exact solution we shall study the simplest possible nonequilibrium quantum system and that is a boundary driven one-dimensional XX spin chain (being equivalent to a tight binding model). It is one of the first driven thermodynamically large quantum systems for which an exact nonequilibrium steady state (NESS) solution has been found~\cite{Karevski:09}, including a compact matrix product form of the NESS~\cite{Znidaric:10b}. Despite being very simple, out of equilibrium it can exhibit interesting phenomena, for instance, when periodically driven its transport properties can vary drastically~\cite{Znidaric:11}. Fluctuations in a closed XX chain, starting from a domain wall initial condition have been studied in Ref.~\cite{Antal:08}. 

We are going to study LD statistics in a NESS that results from the evolution with the Lindblad equation~\cite{Lindblad},
\begin{equation}
\frac{{\rm d}\rho}{{\rm d}t}=\ii [\rho,H]+\sum_{L_j} \cL_0(\rho;L_j)+\cL_{\rm J}(\rho;L_j,0) \equiv \cL(\rho;0),
\label{eq:Lin}
\end{equation}
where we expressed the dissipator as a sum of the ``jump'' term $\cL_{\rm J}(\rho;L_j,s)\equiv 2 {\rm e}^{s} L_j\rho L_j^\dagger$, and the rest $\cL_0(\rho;L_j)\equiv - L_j^\dagger L_j \rho - \rho L_j^\dagger L_j$ (an extra parameter ``$s$'' in $\cL_{\rm J}(\rho;L_j,s)$ will be needed in the LD formalism). The whole Liouvillian propagator is denoted by $\cL(\rho;s=0)$. An especially important state in a NESS $\rho_\infty$ which is a solution of a stationary Lindblad equation, $\cL(\rho_\infty;0)=0$, and to which, in the absence of degeneracies, any initial state converges, $\rho_\infty = \lim_{t \to \infty}\rho(t)$. Expectation value of any observable $G$ in the NESS is $\ave{G}=\tr{(\rho_\infty G)}$. If one wants to calculate fluctuations of $G$ though, or even its complete distribution function, the information contained in $\rho_\infty$ is not enough. Namely, fluctuations in a NESS are connected with nonequilibrium steady-state correlations between $G(t)$ at different times. One way to calculate these is via the LD formalism.

{\em Large deviation formalism.--} The LD formalism has been developed by mathematicians to study properties of distributions beyond the most probable values~\cite{Touchette}. To be concrete, let us consider a current $j$ and its integrated quantity called the number of particles, $N_t=\int_0^t j(\tau)\,{\rm d}\tau$. Provided $N_t$ has a so-called LD property its probability distribution decays exponentially for large $t$ and can be written as $P(N_t\equiv J^{(t)} t) \sim {\rm e}^{-t\,\Phi(J^{(t)})}$, where $\Phi(J)$ is called a LD function and encodes information about the distribution of $N_t$, or, equivalently, of the average current $J^{(t)}\equiv N_t/t$. Provided Lindblad operators are such that the jump term $\cL_{\rm J}$ either changes $N_t$ by $\pm 1$, or leaves it the same, $\Phi$ can be obtained from the largest eigenvalue of the so-called tilted Liouvillian $\cL(\rho;s)$~\cite{Derrida:07,Touchette}. To obtain $\cL(\rho;s)$ we divide Lindblad operators $L_j$ into three sets: the set $S_0$ of those $L_j$ that do not change $N_t$, $S_{+1}$ of $L_j$ that change $N_t$ by $1$, and $S_{-1}$ of operators that change $N_t$ by $-1$, and then form $\cL(\rho;s)=\sum_{j} \cL_0(\rho;L_j)+\sum_{L_j \in S_0}\cL_{\rm J}(\rho;L_j,0)+ \sum_{L_j \in S_{+1}} \cL_{\rm J}(\rho;L_j,s)+\sum_{L_j \in S_{-1}} \cL_{\rm J}(\rho;L_j,-s)$. Denoting by $\lambda(s)$ the eigenvalue of $\cL(\rho;s)$ having the largest real part (we always have $\lambda(0)=0$, with the corresponding eigenvector being $\rho_\infty$), one can show that, provided $\lambda(s)$ is differentiable for all real $s$, it is equal to the scaled cumulant generating function of $N_t$. In other words, $\lambda(s)=\lim_{t \to \infty} \frac{1}{t}\log{\ave{{\rm e}^{s N_t}}}$. Sometimes it is more practical to operate with the distribution function $P(N_t)$ instead of with $\lambda(s)$, which can be obtained from $\lambda(s)$ by the Legendre-Fenchel transform~\cite{foot1},
\begin{equation}
\Phi(J)=\max_{s}\{Js-\lambda(s) \}.
\label{eq:LF}
\end{equation} 
Denoting by $\approx$ the leading contribution in the limit $t \to \infty$, we can write (for $P(J)$ up-to normalization)
\begin{equation}
\ave{{\rm e}^{s N_t}}\approx {\rm e}^{t \lambda(s)},\qquad P(\frac{N_t}{t}=J) \approx {\rm e}^{-t\,\Phi(J)}.
\label{eq:defPhi}
\end{equation}
The scaled cumulants of $N_t$ are given by the derivatives of $\lambda(s)$, and, because the measured current $J^{(t)}$ after time $t$ is simply $J^{(t)}=\frac{N_t}{t}$, they are also equal to $ t^{r-1}\ave{[J^{(t)}]^r}_{\rm c}$,
\begin{equation}
J_r=\frac{1}{t}\langle (N_t)^r \rangle_{\rm c}=t^{r-1}\ave{[J^{(t)}]^r}_{\rm c} = \left. \frac{{\rm d}^r \lambda(s)}{{\rm d}s^r}\right|_{s=0}.
\label{eq:Jr}
\end{equation}
We shall call $J_r$ simply the (scaled) current cumulants. One can see that the LD formalism essentially boils down to the calculation of $\lambda(s)$, more details can be found in, e.g., Ref.~\cite{Touchette}. Formally the LD formalism looks at the outset similar to a more recent full counting statistics~\cite{Levitov}, which also leads to cumulants, and is often employed in mesoscopic physics. With the LD approach though there is an important bonus -- we get a complete distribution function, including statistical physics formalism relating $\Phi(J)$ and $\lambda(s)$. Among many works on full counting statistics we shall only mention Ref.~\cite{Schonhammer:07}, studying a tight binding model with hamiltonian reservoirs, and the recent~\cite{Buca:13}, which uses perturbation theory (in the coupling strength of a boundary Lindblad driving) to derive expansion of current cumulants in spin chains. 

{\em XX spin chain.--}
The Hamiltonian of the XX chain with $n$ sites is $H=\sum_{j=1}^{n-1} \sx_j \sx_{j+1}+\sy_j \sy_{j+1}$, where $\sx,\sy,\sz$ are Pauli matrices. Boundary driving is described by 4 Lindblad operators, $L_1=\sqrt{\Gamma(1+\mu)}\,\sigma^+_1$, $L_2=\sqrt{\Gamma(1-\mu)}\, \sigma^-_1$, $L_3=\sqrt{\Gamma(1-\mu)}\,\sigma^+_n$, and $L_4=\sqrt{\Gamma(1+\mu)}\, \sigma^-_n$. Parameters are the coupling strength $\Gamma$ and the driving strength $\mu$. Such Lindblad operators try to induce magnetization $+\mu$ at the left end and $-\mu$ at the right chain end. For expectation values in $\rho_\infty$ see Ref.~\cite{Znidaric:10b}; here we are interested in statistics of magnetization current~\cite{foot2} (i.e., particle current in fermionic language), defined at site $k$ by $j_k=\sx_k\sy_{k+1}-\sy_k\sx_{k+1}$, that can be determined by counting the amount of transferred magnetization (particles) at right chain end. Therefore, the tilted Liouvillian is obtained by taking $S_0=\{L_1,L_2\}$, $S_{+1}=\{L_4\}$ and $S_{-1}=\{L_3\}$. Using the Jordan-Wigner transformation both $H$ and the Lindblad operators can be written in terms of fermionic operators. The superoperator $\cL(\rho;s)$ is quadratic and thereby its linear action on the space of operators can be as well expressed as a quadratic function of fermionic-like operators acting on the operator space. Such a procedure has been introduced in Ref.~\cite{Prosen:08} and has shown that the eigenvalues of $\cL(\rho;0)$ can be expressed as simple sums over occupation numbers of super-eigenmodes, with the factors being the eigenvalues of the shape matrix. For our driving there is one subtle technical issue: $L_{3,4}$ are not strictly linear in fermions (they contain a string of $\sz_j$). One can show~\cite{Prosen:08} though that for $s=0$ the largest eigenvalue (i.e., the NESS) comes from the even parity sector with an even (zero) number of super-excitations, meaning that the action of a string of $\sz$ gives a trivial phase factor of $+1$. Because $\lambda(s)$ is a smooth function of $s$ we know that we can get it also for $s\neq 0$ by just considering that symmetry sector (there are no degeneracies as a function of $s$). The shape matrix~\cite{Prosen:08} $A$ of size $4n\times 4n$, whose eigenvalues we need, can be calculated and is in the even sector
\begin{equation}
A=\left( \begin{array}{ccccc} B(-\mu,0) & R & 0 & \cdots & 0 \\ -R^{\rm T} & 0 & R & \ddots & 0 \\ 0 & -R^{\rm T} & 0 &  & \vdots \\ \vdots & \ddots &  & \ddots & R \\
0 & 0 & \cdots & -R^{\rm T} & B(\mu,s)
\\ \end{array} \right),
\label{eq:A}
\end{equation}
where $R=\ii\, \sy \otimes \s1$ and 
$B(\mu,s)=\sy \otimes b_1(\mu,s)  + \s1 \otimes b_2(\mu,s)$, with $b_1=-\Gamma \mu\, \sz- \ii\, \Gamma (\mu \cosh{s}+\sinh{s})\, \sx$ and $b_2=-\Gamma(\cosh{s}+\mu\sinh{s})\,\sy$. One can first show~\cite{Supp} that all eigenvalues are at least doubly degenerate and then calculate the characteristic polynomial $q_n$ of a smaller $2n \times 2n$ block tridiagonal matrix, resulting in,
\begin{eqnarray}
q_n(\lambda)&=&(-1)^n\left[ r_n(\lambda)r_n(-\lambda)+a\right],\nonumber \\
a&=&4\Gamma^2\left(\cosh{\frac{s}{2}}+\mu\sinh{\frac{s}{2}}\right)^2-4\Gamma^2,\nonumber \\
r_n(\lambda)&=&\Gamma^2 F_{n-1}(\lambda)+2\Gamma F_n(\lambda)+F_{n+1}(\lambda),
\label{eq:qn}
\end{eqnarray}
where $F_n(x)$ are Fibonacci polynomials. Denoting by $\lambda_j$ the roots of polynomial $q_n(\lambda)$ (\ref{eq:qn}), we can express the largest eigenvalue $\lambda(s)$ of $\cL(\rho;s)$ by a sum over all $\lambda_j$ with positive real parts,
\begin{equation}
\lambda(s)=-4\Gamma+2\sum_{\Re{(\lambda_j)}>0}\lambda_j.
\label{eq:lambda}
\end{equation}

{\em Fluctuation relation.--} The first important observation, which though rather trivially follows from an explicit form of $q_n(\lambda)$ that we derived (\ref{eq:qn}), is that the dependence on $\mu$ and $s$ is only via a constant coefficient $a$. One consequence is that the characteristic polynomial is invariant, i.e., eigenvalues $\lambda_j$ do not change, under the change $s \to -s-c$, where $c=2\ln{\left(\frac{1+\mu}{1-\mu}\right)}$. As a consequence, symmetries $\lambda(s)=\lambda(-s-c)$  as well as $\Phi(J)-\Phi(-J)=-c\,J$ hold, or, equivalently, $P(N_t)/P(-N_t)={\rm e}^{c N_t}$. One can show that the same symmetry holds also in other symmetry sectors and therefore that the whole spectrum of $\cL(\rho;s)$ is the same as the spectrum of $\cL(\rho;-s-c)$~\cite{bojan}. This is nothing but a manifestation of a Gallavotti-Cohen-like symmetry or a nonequilibrium fluctuation theorem~\cite{GC}. As we shall see, the divergence of $c$ at $\mu=1$ will be reflected in a discontinuous change of the definition range of current $J$.

{\em Current cumulants.--}
\begin{figure}[ht!]
\centerline{\includegraphics[width=0.45\textwidth]{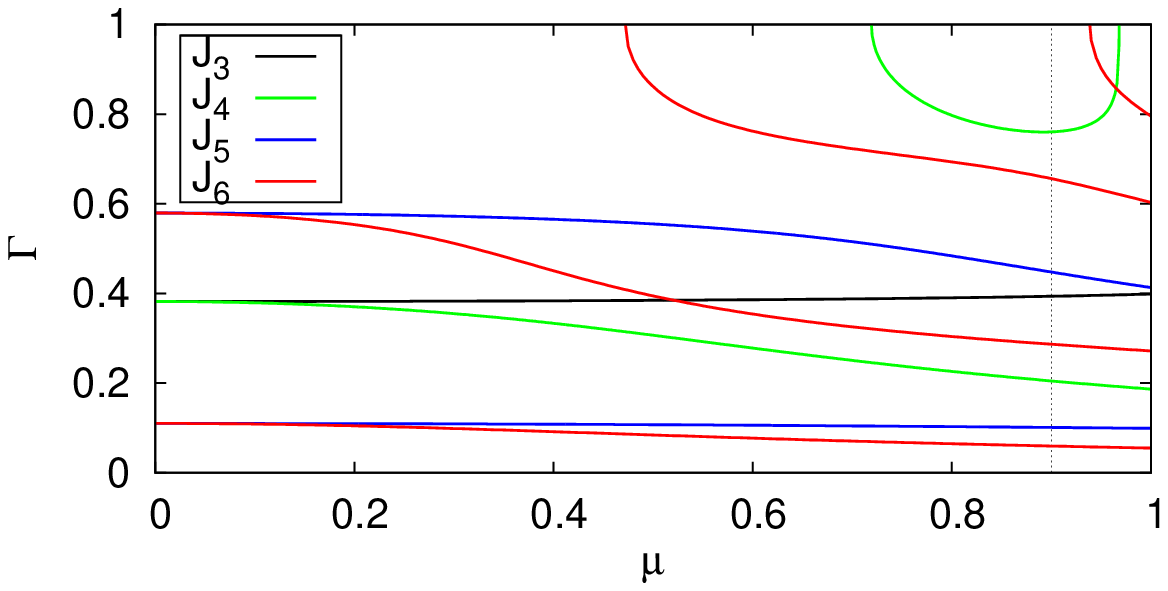}}
\centerline{\includegraphics[width=0.45\textwidth]{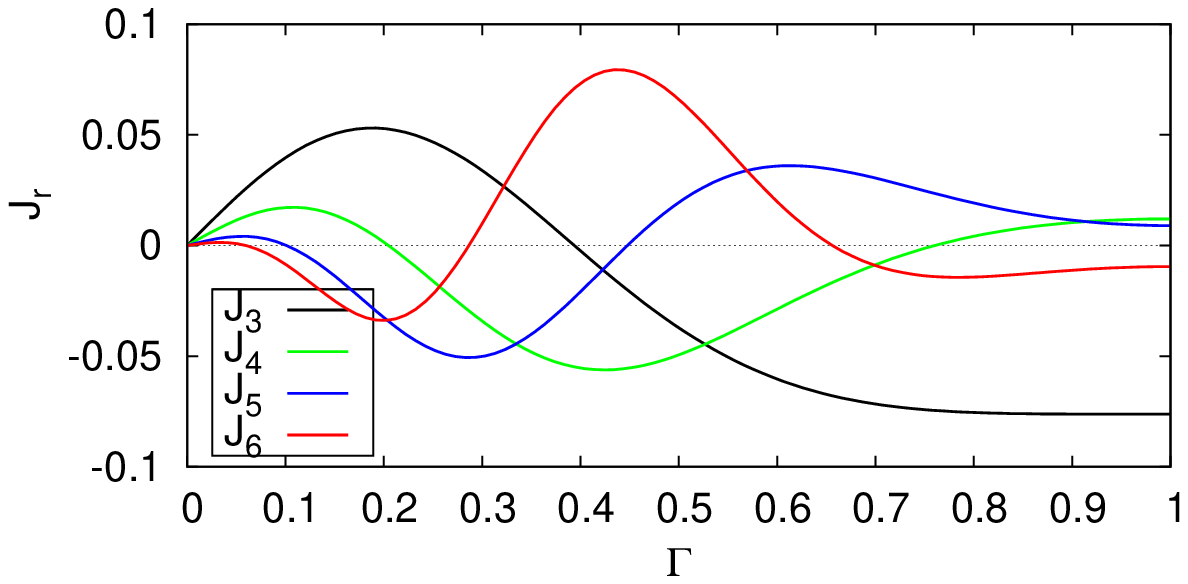}}
\caption{Top: Location of zeros (curves) of the first few current cumulants $J_r$. Bottom: Dependence of $J_r$ on $\Gamma$ for $\mu=0.9$ (cross-section along the dashed line in the upper plot).}
\label{fig:nicle}
\end{figure}
Current cumulants (\ref{eq:Jr}) are obtained from Eq.~(\ref{eq:lambda}). We therefore need derivatives of roots $\lambda_j$ of $q_n$ at $s=0$. These can in turn be obtained by taking derivatives over $s$ of the characteristic polynomial $q_n(\lambda)$. Doing that we obtain
\begin{equation}
J_1=2\sum_{\Re{(\lambda_j^0)}>0} A_n(\lambda_j^0),\quad A_n(x)\equiv \frac{-a'_0}{r'_n(x)r_n(-x)},
\label{eq:J1}
\end{equation}
where $\lambda_j^0\equiv\lambda_j(s=0)$ (note that $a(s=0)=0$), $a'_0\equiv a'(s=0)={\rm d}a/{\rm d}s=4\Gamma^2\mu$ and $r'_n(\lambda)$ is a derivative of $r_n(\lambda)$ (\ref{eq:qn}) with respect to $\lambda$. For higher cumulants expressions get more complicated; we will write out just the one for $J_2$,
\begin{equation}
J_2=\frac{a''_0}{a'_0}J_1-2\sum_{\Re{(\lambda_j^0)}>0} A_n^2(\lambda_j^0)\left( \frac{r''_n(\lambda_j^0)}{r'_n(\lambda^0_j)}-2\frac{r'_n(-\lambda^0_j)}{r_n(-\lambda^0_j)}\right).
\label{eq:J2}
\end{equation}
For small $n$ one can explicitly express zeros $\lambda_j^0$ and therefore also cumulants in terms of parameters $\mu$ and $\Gamma$. Importantly, one notes that $J_r$ is independent of $n$ provided $n>r$~\cite{foot3}! Therefore, one can obtain $J_r$ in the TDL already by calculating derivatives of $\lambda(s)$ for $n=r+1$, enabling one to get exact expressions for lower cumulants,
\begin{eqnarray}
J_1&=&\frac{2\mu}{\epsilon},\quad J_2=\frac{\epsilon^2-3\mu^2}{\epsilon^3},\nonumber \\
J_3&=&\frac{\mu}{2\epsilon^5}\left[\epsilon^2(\epsilon^2-9)-3\mu^2(\epsilon^2-10) \right],
\label{eq:J123}
\end{eqnarray}
where they are expressed as functions of $\epsilon \equiv \Gamma + \frac{1}{\Gamma}$. Exact expressions for $J_{4,5,6}$ in the TDL can be found in~\cite{Supp}. The general form of $J_r$ is $J_r=\frac{\mu^\alpha p_r(\mu,\epsilon)}{2^{r-1}\epsilon^{2r-1}}$, where $\alpha=1$ for odd $r$ while $\alpha=0$ for even $r$, and $p_r$ is an even polynomial of degree $4$ in $\mu$ and of degree $2(r-1)$ in $\epsilon$. An important observation is also that all cumulants are functions of only $\epsilon$, meaning that $J_r$, LD function $\Phi(J)$ as well as $\lambda(s)$ are, in the TDL, invariant under mapping $\Gamma \to \frac{1}{\Gamma}$ (note that this invariance does not hold for some other observables in $\rho_\infty$, for instance for magnetization~\cite{Znidaric:10b}). Observe that $J_r$ have, apart from $J_1$, a rather nontrivial dependence on $\mu$ and $\Gamma$, see Fig.~\ref{fig:nicle}.

\begin{figure}[ht!]
\centerline{\includegraphics[width=0.45\textwidth]{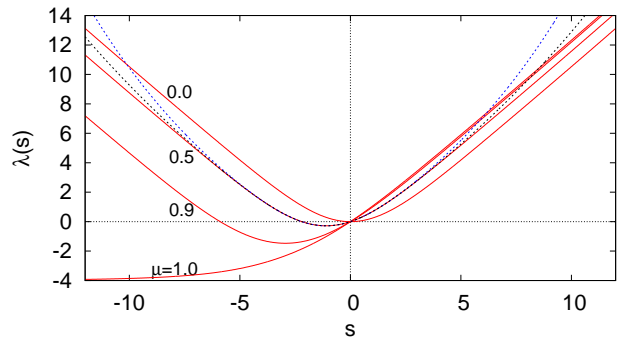}}
\caption{Cumulant generating function $\lambda(s)$ in the thermodynamic limit (full red lines) for different values of driving $\mu$ ($\Gamma=1$). The asymptotic linear behavior of $\lambda(s)$ reflects a strict upper and lower bound on the possible current values (\ref{eq:LF}). Two dotted lines for $\mu=0.5$ are finite-$n$ results (blue for $n=4$ and black for $n=6$).}
\label{fig:lambda}
\end{figure}

\begin{figure}[ht!]
\centerline{\includegraphics[width=0.45\textwidth]{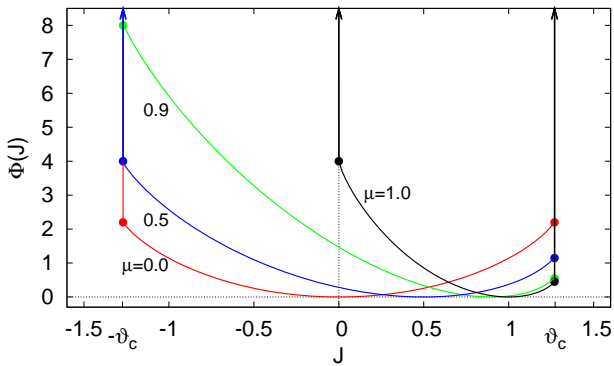}}
\caption{Large deviation rate function $\Phi(J)$ in the thermodynamic limit ($\Gamma=1$). $\Phi(J)$ is a non-smooth function of $J$. The definition range discontinuously changes from $(-\vartheta_{\rm c},\vartheta_{\rm c})$ for $\mu<1$ to $(0,\vartheta_{\rm c})$ at $\mu=1$.}
\label{fig:phi}
\end{figure}

{\em Large deviation function.--}
We were not able to obtain an analytic expression for $\lambda(s)$ in the TDL and for large $s$. However, zeros of $q_n(\lambda)$, and therefore also $\lambda(s)$, can be computed numerically for sizes $n$ of several hundred. For not too large $s$ the convergence with $n$ is actually quite fast (see Fig.~\ref{fig:lambda}).   
An important property of $\lambda(s)$ is that for large $|s|$ the derivative $\lambda'(s)$ converges to a constant. A consequence of that is that using the Legendre-Fenchel transform (\ref{eq:LF}) results in $\Phi(J)$ being infinite for $J$ that are larger/smaller than the minimal/maximal slope of $\lambda(s)$. This can be nicely seen in Fig.~\ref{fig:phi} where we show $\Phi(J)$ in the TDL. Remember that the zeros of $\Phi(J)$ correspond to the most probable values (in our case the NESS expectation value $J_1$).
Several interesting things can be observed. First, $\Phi(J)$ is finite, and therefore also the distribution function of the current, only within the interval $(-\vartheta_{\rm c},\vartheta_{\rm c})$ for all $\mu<1$ and within $(0,\vartheta_{\rm c})$ for $\mu=1$. Definition range of the current discontinuously changes at $\mu=1$. One can calculate~\cite{Supp} that $\vartheta_{\rm c}=\frac{4}{\pi}$ and is independent of $\mu$ and $\Gamma$. Overall, $\Phi(J)$ and $\lambda(s)$ have an inessential dependence on $\Gamma$. At the boundaries of this definition range $\Phi(J)$ has an infinite jump. This is in turn reflected in a discontinuity of the distribution function $P(J)$, for instance, $P(\vartheta_{\rm c}-0)$ is finite while $P(\vartheta_{\rm c}+0)=0$. A nonanalytic behavior of a thermodynamic potential is a characteristic feature of phase transitions and one can say that the open XX chain is a model residing at a phase transition point. This explains and confirms a little mysterious finding in Ref.~\cite{Znidaric:10} that the XX model exhibits a discontinuous change from being a ballistic conductor to being a diffusive conductor as one adds nonzero dephasing. Just from $\rho_\infty$, studied there, it was not clear where that discontinuity came from.
\begin{figure}[ht!]
\centerline{\includegraphics[width=0.45\textwidth]{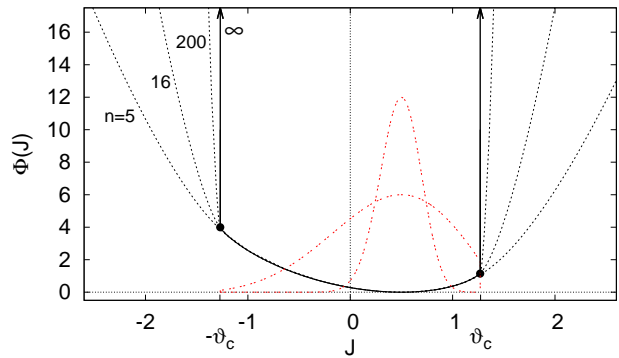}}
\caption{Convergence of large deviation rate function with system size $n$ ($\mu=0.5$, $\Gamma=1$). Dashed red lines are unnormalized distributions $\sim \exp{(-t\,\Phi(J))}$ (shown for $t=1$ and $t=10$, and $n\to\infty$), trying to indicate how the current distribution looks.}
\label{fig:converg}
\end{figure}
It is actually instructive also to look at finite-$n$ values of $\lambda(s)$ and $\Phi(J)$, see Fig.~\ref{fig:converg} and~\ref{fig:lambda}. One observes that for any finite $n$, at a sufficiently large $|s|$, $\lambda(s)$ eventually begins to increase/decrease faster than with the slope $\vartheta_{\rm c}$. This means that for finite $n$ thermodynamic functions are (expectedly) smooth and that the definition range of the current is the whole real axis for $\mu<1$, or the whole positive real axis for $\mu=1$.

{\em Conclusion.--}
We present results for the large deviation statistics of a nonequilibrium quantum XX spin chain driven at the boundaries. Despite being a simple textbook model (a driven tight binding model), we find a rather rich behavior out of equilibrium. In the thermodynamic limit the large deviation function exhibits nonanalyticity, signaling a phase-transition-like behavior, reflected for instance in a discontinuity of the current distribution function. For a few lower-order cumulants we provide exact closed expressions. We also explicitly show the validity of a nonequilibrium fluctuation theorem. The study elucidates the importance of large deviation formalism for quantum systems, since, for instance, the nonanalytic properties discovered here are not visible in simple steady-state expectation values.\\

\section*{Appendix}

\subsection{Characteristic polynomial}
We need eigenvalues of the shape matrix $A$ (\ref{eq:A}) of dimension $4n\times 4n$. $A$ is block tridiagonal with blocks of size $4$. We first observe~\cite{Prosen:10} that in $R$ as well as in $B$ the matrix on the 1st tensorial subspace is either $\sy$ or the identity. Therefore, by simple $2$ dimensional rotation $A$ can be brought to an upper triangular form, showing that all eigenvalues of $A$ are at least twice degenerate and that all eigenvalues can be obtained from smaller $2n\times 2n$ matrix $\tilde{A}$ (each eigenvalue of $\tilde{A}$ appears twice in the spectrum of $A$),
\begin{equation}
\tilde{A}=R_1 \otimes \s1+D_1\otimes B_1+D_n \otimes B_n,
\end{equation}
where $n \times n$ matrices are $[D_r]_{jk}=\delta_{j,r}\delta_{k,r}$, $[R_1]_{jk}=\ii(\delta_{j-1,k}+\delta_{j+1,k})$, while $B_1=b_1(-\mu,0)+b_2(-\mu,0)$ and $B_n=b_1(\mu,s)+b_2(\mu,s)$. $\tilde{A}$ is again block-tridiagonal. Characteristic polynomial of a block-tridiagonal matrix can be written in terms of transfer matrices~\cite{Molinari:08}. Defining a $4\times 4$ matrix $T$ as
\begin{equation}
T=
\begin{pmatrix}
-B^\lambda_n & -\ii \mathbbm{1}_2\\
\mathbbm{1}_2 & 0
\end{pmatrix}
\begin{pmatrix}
-\ii \lambda \mathbbm{1}_2 & -\mathbbm{1}\\
\mathbbm{1}_2 & 0
\end{pmatrix}
^{n-2} 
\begin{pmatrix}
\ii B^\lambda_1 & \ii \mathbbm{1}_2\\
\mathbbm{1}_2 & 0
\end{pmatrix},
\end{equation}
where $B_k^\lambda=B_k-\lambda \mathbbm{1}_2$, the determinant of $\tilde{A}-\lambda\mathbbm{1}$ is equal to $(-1)^{n-1}$ times the determinant of the upper-left $2\times 2$ block of $T$, giving
\begin{eqnarray}
q_n(\lambda) &\equiv& \det{[\tilde{A}-\lambda \mathbbm{1}_{2n}]}=(-1)^{n-1}{\rm det}\,[ -\ii p_{n-2}(B_n^\lambda B_1^\lambda)+\nonumber \\
&+&p_{n-3}(B_n^\lambda+B_1^\lambda)+\ii p_{n-4}\mathbbm{1}_2 ],
\end{eqnarray}
where $p_k$ are polynomials in $\lambda$ given by the recursion $p_{k+1}=-\ii\lambda p_k-p_{k-1}$, with starting points $p_{-1}=0, p_0=1$. Recursion can be solved, getting $p_k={\rm e}^{-\ii \frac{\pi}{2}k}\lambda^m\sum_{j=0}^{d}\lambda^{2j} {d+m+j \choose d-j}$, where $d\equiv \lfloor \frac{k}{2}\rfloor$ is the divisor and $m\equiv k-2\tilde{k}$ the remainder. Up to a prefactor these are in fact Fibonacci polynomials $F_{n+1}(\lambda)$, and we have $p_k(\lambda)={\rm e}^{-\ii \frac{\pi}{2}k} F_{k+1}(\lambda)$. Characteristic polynomial can therefore be finally written as
\begin{eqnarray}
q_n(\lambda)&=&(-1)^n\left[ r_n(\lambda)r_n(-\lambda)+a\right],\nonumber \\
a&=&4\Gamma^2\left(\cosh{\frac{s}{2}}+\mu\sinh{\frac{s}{2}}\right)^2-4\Gamma^2,\nonumber \\
r_n(\lambda)&=&\Gamma^2 F_{n-1}(\lambda)+2\Gamma F_n(\lambda)+F_{n+1}(\lambda).
\end{eqnarray}
Fibonacci polynomials $F_n(x)$ satisfy recursion $F_{n+1}=xF_n+F_{n-1}$, with starting $F_1=1$, $F_2=x$. $F_n(x)$ is of order $n-1$ in $x$. Because all coefficients of $q_n(\lambda)$ are positive complex roots come in conjugate pairs $\lambda_j$, $\lambda_j^*$, and because $q_n$ is even in $\lambda$ all roots come in pairs $\lambda_j,-\lambda_j$, see also illustration in Fig.~\ref{fig:roots}.

\subsection{Calculation of the extreme current $\vartheta_{\rm c}$}
Critical current $\vartheta_{\rm c}$ is equal to the asymptotic slope of $\lambda(s)$ (in the limit $n \to \infty$ and $s\to \infty$). Instead of directly calculating the derivative of $\lambda(s)$ we can evaluate it in a simpler way as $\lambda(s\to \infty)/s$,
\begin{equation}
\vartheta_{\rm c}=2\lim_{s\to\infty} \lim_{n\to \infty}\frac{1}{s}\sum_{\Re{(\lambda_j)}>0}\lambda_j(s),
\label{eq:Jc}
\end{equation}
with $\lambda_j(s)$ as always being the roots of $q_n$ (\ref{eq:qn}). Beware that the order of the two limits is crucial -- exchanging them gives infinity. As we shall see, at fixed $s$ the roots $\lambda_j(s)$, as one increases $n$, become increasingly close to those in the case $\Gamma=0$. The idea then is to use perturbation theory to obtain $\lambda_j(s)$ in the TDL.

Before tackling that problem let us have a look at a simpler case of perturbation theory at $s=0$ for which $a=0$. Setting also $\Gamma=0$ the situation is trivial, with the characteristic polynomial being $q_n(\Gamma=0)=F_{n+1}^2(\lambda)$, where we used that $F_{n+1}(x)$ are even/odd functions of $x$ for odd/even $n+1$. Roots of $F_n$ are just $-2\ii \cos{(\pi j/n)}, j=1,\ldots,n-1$, and we get that $\lambda_j(\Gamma=0)=-2\ii \cos{(\pi t_j)}$, with $t_j=j/(n+1), j=1,\ldots,n$ (every $\lambda_j$ being a double root of $q_n$). These are of course nothing but the eigenvalues of an uncoupled tight binding model (i.e., of a harmonic chain). For small $\Gamma$ one can use perturbation theory. Doing that we have $\Delta\lambda_j=-q_n(\lambda_j)/q'_n(\lambda_j)$. Using that $F_n'=(2nF_{n-1}+(n-1)xF_n)/(4+x^2)$, we get in the lowest order in $\Gamma$ the expression $\lambda_j=-2\ii \cos{(\pi t_j)}+\frac{4\Gamma}{n+1}\sin^2{(\pi t_j)}$, with $t_j=j/(n+1), j=1,\ldots,n$. Important here is a prefactor $\sim 1/n$, comming from $F_n'$, and causing the correction to become small in the TDL. Incidentally, from perturbative expression for $\lambda_j$ we can also read out the value of the gap $\Delta$ of the Liouvillian in the even subsector, $\Delta = \frac{16\Gamma}{n+1} \sin^2{\frac{\pi}{(n+1)}}\asymp 16\pi^2\Gamma/(n+1)^3$. Note that this expression is not necessarily exactly equal to the gap of the whole spin Liouvillian $\cL(\rho;0)$ because of the symmetry-subspace complications connected with a string of $\sz$ in $L_{3,4}$.

\begin{figure}[th!]
\centerline{\includegraphics[width=0.44\textwidth]{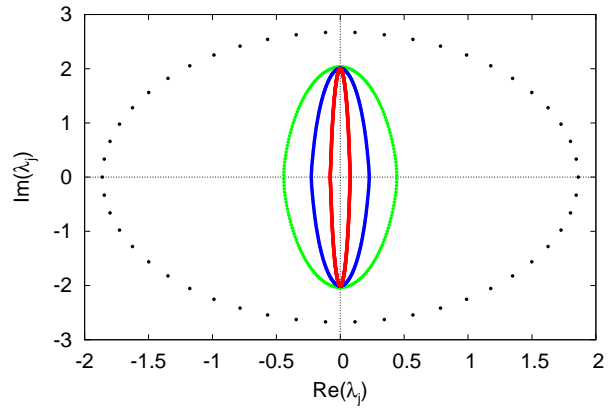}}
\caption{Location of roots of $q_n(\lambda)$ for $s=40$ ($\Gamma=1, \mu=\frac{1}{2}$). Shown are cases for $n=25, 100, 200$ and $600$ (different colors, from largest to smallest ``ellipse'').}
\label{fig:roots}
\end{figure}

For nonzero $s$ and large $n$ we can use the same reasoning. Roots become closer and closer to the imaginary axis (because of an $\sim 1/n$ prefactor in the correction), so that eventually one can use perturbation theory, see Fig.~\ref{fig:roots}. To evaluate $\vartheta_{\rm c}$ (\ref{eq:Jc}) we also only need real parts of $\lambda_j$ (due to complex-conjugate pairs of roots imaginary parts cancel). For those we get (fixing for simplicity $\Gamma=1$, $\mu=1$) $\Re{(\lambda_j)}\asymp \frac{s}{n}\sin{(\pi t)}$, $t \in [0,1]$. Replacing summation in (\ref{eq:Jc}) by an integral we finally get $\vartheta_{\rm c}=\frac{2}{s}s\int_0^1 \sin{(\pi t)}{\rm d}t=\frac{4}{\pi}$. From Fig.~\ref{fig:phi} one can see that this value nicely agrees with the location of the divergence in $\Phi(J)$.

\subsection{Higher-order cumulants}
The procedure to calculate higher order cumulants is analogous to calculation of $J_{1,2,3}$. Intermediate steps are a bit messy and we shall write only final results,
\begin{eqnarray}
J_4
&=&\frac{1}{4\epsilon^7}\left[ \epsilon^4(\epsilon^2-9)-30\mu^2\epsilon^2(\epsilon^2-6)+15\mu^4(6\epsilon^2-35)\right],\nonumber\\
J_5&=&\frac{\mu}{8\epsilon^9}\left[\epsilon^4 d_4-30\mu^2\epsilon^2 d_2+90\mu^4d_0 \right],
\end{eqnarray}
where $d_4=\epsilon^4-90\epsilon^2+450$, $d_2=(\epsilon^2-5)(\epsilon^2-35)$, and $d_0=\epsilon^4-35\epsilon^2+147$ (remember that $\epsilon\equiv \Gamma+\frac{1}{\Gamma}$). For $J_6$ one gets
\begin{equation}
J_6=\frac{1}{16\epsilon^{11}}\left[\epsilon^6 f_6-21\mu^2\epsilon^4f_4+\mu^4\epsilon^2f_2-\mu^6f_0 \right],
\end{equation}
where $f_6=\epsilon^4-90\epsilon^2+450$, $f_4=13\epsilon^4-300\epsilon^2+1125$, $f_2=1050(3\epsilon^4-55\epsilon^2+189)$, and $f_0=315(25\epsilon^4-420\epsilon^2+1386)$. One could also calculate few even higher cumulants, however, already $J_6$ is rather complicated and such exact expressions, per se, do not offer much physical insight.  

An interesting observation that we made is that $J_r$ is independent of $n$ for $n>r$. This is a consequence of various identities involving roots of linear combinations of Fibonacci polynomials. The simplest one, needed in $J_1$, states that, if $r_n(x)=a F_{n-1}(x)+b F_n(x)+F_{n+1}(x)$ ($a,b$ are arbitrary complex numbers) and $x_j$ are solutions of $r_n(x_j)=0$, then
\begin{equation}
\sum_{x_j} \frac{1}{r'_n(x_j)r_n(-x_j)}=\frac{1}{2(1+a)b},
\end{equation}
for any $n>1$.


\begin{thebibliography}{}
\bibitem{foot1} When $\lambda(s)$ is strictly convex the transformation reduces to the standard Legendre transform, $\Phi(J)=Js_*-\lambda(s_*)$, with $s_*$ being the unique solution of ${\rm d}\lambda(s_*)/{\rm d}s=J$.
\bibitem{foot2} Note that, because we use ${\rm e}^s$ instead of ${\rm e}^{2s}$ in our definition of $\cL(\rho;s)$, we are in fact studying the particle current and not true magnetization current, being equal to $2j_k$. The central moment $\tilde{J}_r$ of the magnetization current can be simply obtained as $\tilde{J}_r=2^r J_r$.
\bibitem{foot3} This gives an amusing new relation saying that particular sums involving zeros of linear combinations of Fibonacci polynomials from $r_n$ are independent of $n$, see~\cite{Supp}.
\end{thebibliography}
\end{document}